\documentclass[aps,prl,twocolumn,groupedaddress,showpacs]{revtex4}

\usepackage{graphicx}

\begin{document}


\title{Self-Organization of Vortex Length Distribution in Quantum Turbulence:\\ An Approach from the Barab\'asi-Albert Model}


\author{Akira Mitani and Makoto Tsubota}
\affiliation{Department of Physics, Osaka City University, Sumiyoshi-ku, Osaka 558-8585, Japan}


\date{\today}

\begin{abstract}
The energy spectrum of quantum turbulence obeys Kolmogorov's law. The vortex length distribution (VLD), meaning the size distribution of the vortices, in Kolmogorov quantum turbulence also obeys a power law. We propose here an innovative idea to study the origin of the power law of the VLD. The nature of quantized vortices allows one to describe the decay of quantum turbulence with a simple model that is similar to the Barab\'asi-Albert model of large networks. We show here that such a model can reproduce the power law of the VLD fairly well.
\end{abstract}


\maketitle
Scale-free (scale invariant) systems are ubiquitous. Examples of scale-free systems include the World Wide Web (WWW), whose vertices are HTML documents connected by hyperlinks pointing from one page to another, and social networks such as a movie collaboration network of movie actors in which each actor is represented by a vertex, two actors being connected if they were cast together in a same movie. Regardless of the system and the identity of its constituents, the probability $P(k)$ that a vertex in the network connects with other $k$ vertices decays as a power law, following $P(k)\propto k^{-\gamma}$ \cite{Albert2002}.

Fluid turbulence has a similar scaling property. Theoretical and experimental studies have shown that classical turbulence has the scale invariance in its energy spectrum $E(k)\equiv 4\pi k^2|\tilde{v}(k)|^2$, where $\tilde{v}(k)$ is the Fourier transform of fluid velocity $v(\boldmath{r})$ \cite{Frisch-text}. This energy spectrum obeys the Kolmogorov power law $E(k)\propto k^{-5/3}$, which means that classical turbulence has scaling property in wave number space. Some numerical simulations recently revealed that decaying quantum turbulence at zero temperature, in which quantized vortex lines of various length form an irregular tangle \cite{Donnelly-text}, is also a scale invariant system \cite{Araki2002, Kobayashi2005}; quantized vortices are topological defects of order parameter in a superfluid. These authors calculated the energy spectrum of the velocity field and found that it obeys the Kolmogorov power law $E(k)\propto k^{-5/3}$ as in classical turbulence. The Kolmogorov law is a scaling property in wave number space, but should be closely related with the self-similarity of the turbulent velocity field in real space. So we ask the following question: how can the scale invariance in wave number space be translated into the scale invariance in real space? A possible answer was proposed by Araki \textit{et al.} \cite{Araki2002}. They calculated the vortex length distribution (VLD) $n(l)$, where $n(l)\Delta l$ represents the number of vortices in Kolmogorov quantum turbulence of length from $l$ to $l+\Delta l$. This VLD obeys a scaling property $n(l) \propto l^{-\alpha}$ with $\alpha=1.34\pm 0.18$. Kobayashi also found a scaling property of $n(l)$ with $\alpha \approx 1.5$ \cite{Kobayashi-private}.

Why can such diverse systems acquire scale invariant properties? In the case of the large networks such as WWW, Barab\'asi and Albert showed that the scaling property comes from two generic mechanisms: (i) networks expand continuously by the addition of new vertices, and (ii) new vertices are more likely to attach to sites that have more connections than those with fewer connections \cite{Barabasi1999}. While, there has been no practical study that answers this question about the scale invariance in quantum turbulence from a generic point of view \cite{explain3}, although many researchers believe that this scaling property results from the Richardson cascade process, through which large vortex loops continuously split into smaller loops. The aim of this paper is to propose an innovative method to study the origin of the scaling property in quantum turbulence at zero temperature using the Barab\'asi-Albert model, hereafter the BA-model.

We now describe the BA-model for large networks. Starting with a small number ($m_0$) of vertices, at every time step they added a new vertex with $m(\leq m_0)$ edges that link the new vertex to $m$ different vertices already present in the system. The probability $\Pi$ that a new vertex will be connected to a vertex $i$ was assumed to depend linearly on the connectivity $k_i$ of that vertex, so that $\Pi(k_i)=k_i/\Sigma_j k_j$. After $t$ time steps, the model leads to a random network with $t+m_0$ vertices and $mt$ edges. They calculated the probability $P(k)$ by numerical and analytical methods. Both calculations suggested that the probability $P(k)$ of this model is proportional to $k^{-3}$, which is approximately consistent with the power laws observed in various large networks.

Before applying the above model to quantum turbulence, we briefly review the nature of quantized vortices at zero temperature. First, quantum turbulence can be treated as an assembly of vortex loops of various sizes when the system is so large that the boundaries have negligible influence on the system. Quantized vortices are stable, and thus the only ways that the topology of vortex loops can change are vortex-vortex reconnections. Second, although vortex-vortex reconnections can both divide one vortex loop into two (the split type) and combine two vortex loops into one (the combination type) (see Fig. 12 in \cite{Tsubota2000}), occurrence frequency of the split type reconnection is so larger than that of the combination type that we can neglect the latter type. Actually, this is confirmed numerically for a dilute vortex tangle at zero temperature \cite{Tsubota2000}. Third, the total vortex line length is nearly conserved during the dynamics, even when vortex reconnections occur. The only exceptional event is annihilation of small-scale vortices. Reduction of vortex line length may also occur due to energy dissipation (e.g., phonon emission) caused by the Kelvin wave cascade \cite{Vinen2003} and also by vortex reconnections \cite{Leadbeater2001}. However, the reductions in these cases are small relative to the overall vortex dynamics that we can neglect the reduction of line length \cite{Leadbeater2001, Leadbeater2003}. This fact permits us to reach another important conclusion: VLD does not change without reconnections.  Finally, according to the dynamical scaling law of quantized vortices \cite{Schwarz1988}, a vortex loop with a configuration that is similar to that of another, except that its size is reduced by a factor $\lambda (<1)$, will follow a similar motion as the larger loop except the time scale is shortened by $\lambda^2$.  Hence, we can make the following assumption. If the scale of the vortex loop 'A' is $\lambda$ times smaller than another vortex loop 'B', the probability that the vortex loop 'A' reconnects with itself per unit time is $(1/\lambda)^2$ times larger than that of 'B'.

We confirmed the validity of this assumption by direct numerical simulation of vortex dynamics under the vortex filament model as follows. The method of the numerical simulation is similar to that of previous study \cite{Tsubota2000}, in which we followed the dynamics of vortex filaments represented by single strings of points. However, we adopt here a new algorithm of vortex-vortex reconnection process, based on considerations of crossing lines \cite{Kondaurova05, explain1}. We prepared a random polygon as a tentative initial vortex configuration which follows a probability function \cite{Cloizeaux1979}
\begin{eqnarray}
P(\mathbf{u}_1,\cdots ,\mathbf{u}_N)=Z^{-1}\delta(\mathbf{u}_1+\cdots+\mathbf{u}_N) \nonumber \\ \times\mathrm{exp}\left(-\sum^{N}_{j=1} \frac{\mathbf{u}_j^2}{2(sz)^2}\right).
\label{eq:probab}
\end{eqnarray}
Here $N$ is the maximum number of points on the initial configuration, $\mathbf{u}_j$ is the vector from the position of the $(j-1)$-th point on a vortex filament to that of the $j$-th point, $\delta(x)$ is the $\delta$-function, $z$ is spatial resolution of our simulation, $s$ is a scale factor, and the normalization factor is given by
\begin{eqnarray}
Z=\int d\mathbf{u}_1 \cdots \int d\mathbf{u}_N \delta(\mathbf{u}_1+\cdots+\mathbf{u}_N)\nonumber \\ \times\mathrm{exp}\left(-\sum^{N}_{j=1} \frac{\mathbf{u}_j^2}{2(sz)^2}\right).
\end{eqnarray}
Because $sz$ corresponds to an average distance between neighboring points, we can change the scale of the initial vortex loop by controlling $s$ \cite{explain2}. When $s>1$, this tentative configuration has larger distances between points than the spatial resolution of the simulation. Hence, we can complete the initial configuration by interpolating points between the tentative points by using an adaptive meshing routine \cite{Tsubota2000}. We follow the dynamics from the initial configuration and obtain the time $\tau_s$ when the vortex makes the first self-reconnection for $s=5$ and $s=10$. Although the time $\tau_s$ is sensitive to the initial configuration, we see that the quantity $\langle \tau_{10}\rangle /\langle \tau_5\rangle$ becomes $3.758\approx 4=2^2$ (i.e., nearly the square of the ratio of the size scales), where the brackets represents ensemble average of 1,000 samples. This result justifies our assumption about the dependence of occurrence frequency of reconnections on the scale of vortex loops.

\begin{figure}
\begin{center}
\includegraphics[width=.7\linewidth]{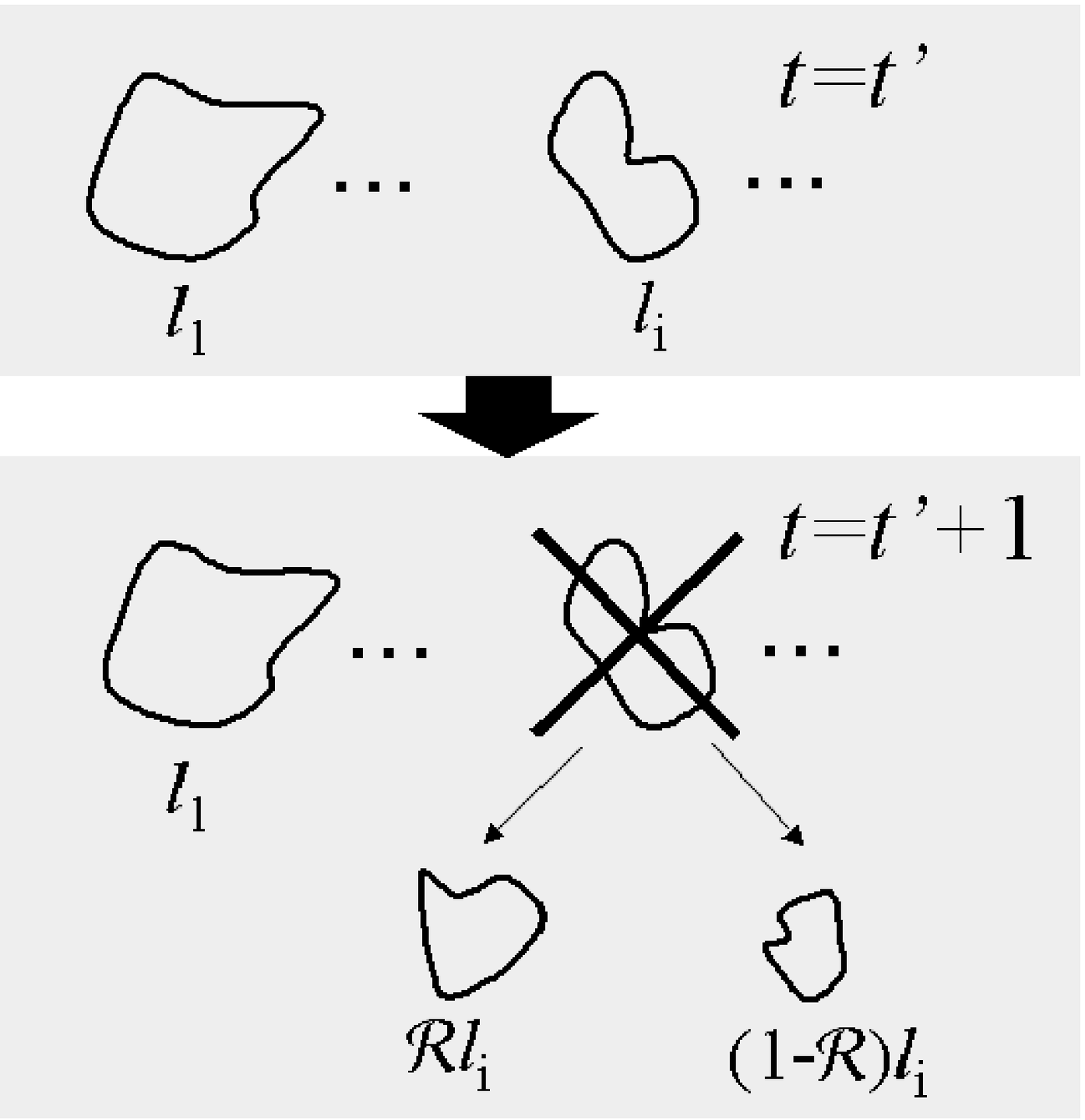}
\caption{Schematic of the model. At every time step one, and only one, loop is randomly chosen to divide into two daughter loops with the probability given by Eq. (\ref{eq:dep}). If the chosen loop has length $l_i$, $\mathcal{R}l_i$ and $(1-\mathcal{R})l_i$ are assigned as the length of these two loops.}
\label{fig:sche}
\includegraphics[width=.8\linewidth]{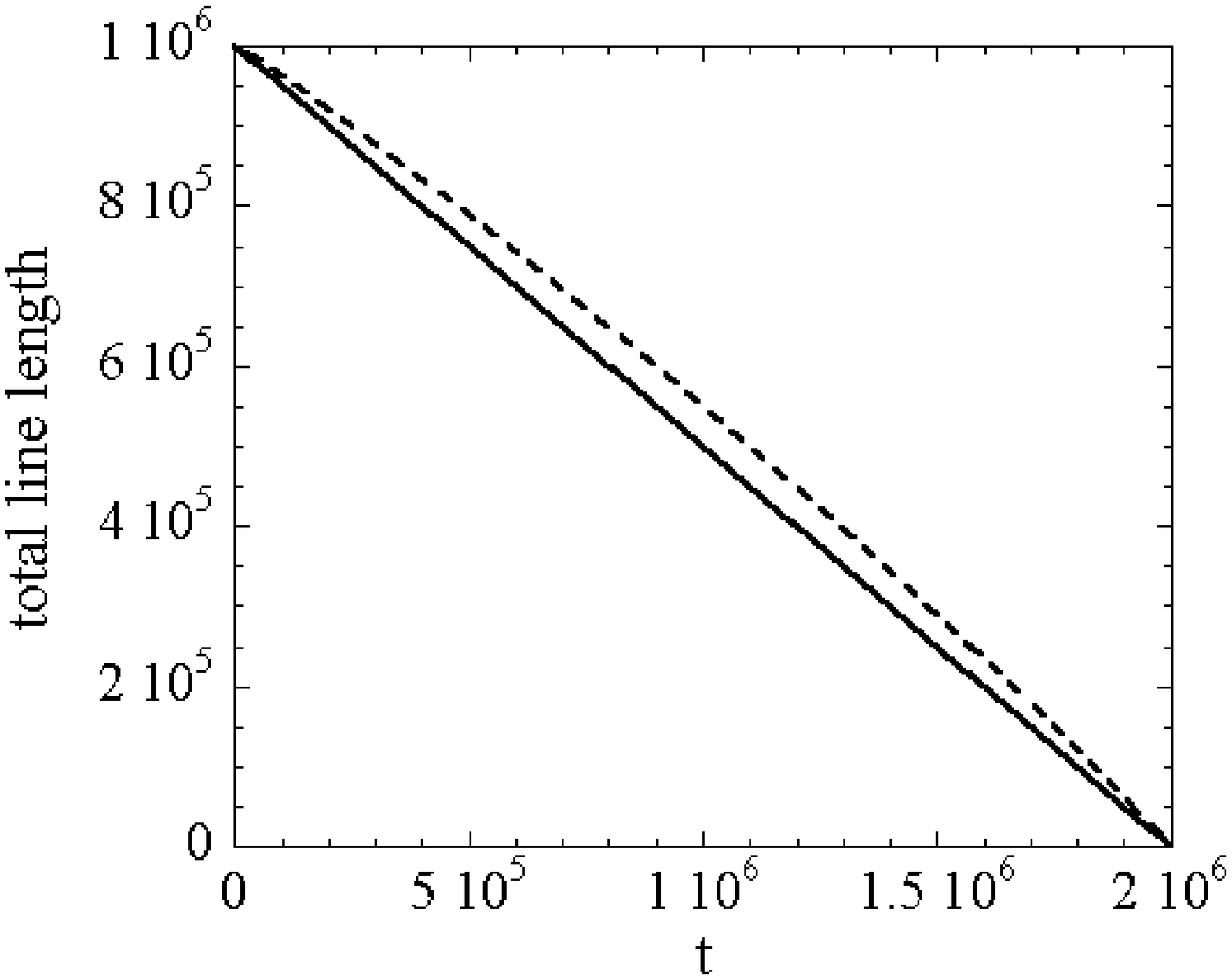}
\caption{Solid and slashed lines represent time dependences of total line length assuming $\Pi(l)$ as Eq.(\ref{eq:dep}) and eq (\ref{eq:const}), respectively.}
\label{fig:total_line}
\end{center}
\end{figure}

The facts listed above permit us to propose the following reasonable hypothesis. \textit{The observed power law of the VLD can be simply understood as a consequence of continuous reconnections, in which the occurrence frequency is governed statistically by the dynamical scaling law of quantized vortices.} In this paper, to clarify the role of Richardson cascade process, we focus on dilute quantum turbulence, in which split type reconnections are dominant and correlation of a vortex loop with other loops can be negligible. We now show that the decaying process of dilute quantum turbulence at zero temperature can be described using a simple model similar to the BA-model as the following. It must be noted that the system we discuss here is different from that concerned by the original BA-model. Hence, the two essential points of BA-model, increasing the number of the vertices and inequality among the vertices, appear in different ways in our model. In the case of decaying process of superfluid turbulence, the number of constituents increases by splitting of vortex loops in the system. The inequality among the vertices getting linked corresponds to the inequality of getting split, which comes from dynamical scaling law, in our case of decaying process of superfluid turbulence. Starting with a small number ($m_0$) of loops, each of which have their own length $l_i$, at every time step we split one of them into two daughter loops. When a loop of length $l$ is divided into two, the daughter loops have lengths $\mathcal{R}l$ and $(1-\mathcal{R})l$, where $\mathcal{R}$ is a random number ($0<\mathcal{R}<1$) (Fig.\ref{fig:sche}). To incorporate the scaling effect on the occurrence frequency governed by the dynamical scaling law, we assume that the probability $\Pi$ that the $i$-th loop are chosen to divide into two depends on its length $l_i$, so that
\begin{equation}
\Pi(l_i)=\frac{(1/l_i)^2}{\Sigma_j (1/l_j)^2}.
\label{eq:dep}
\end{equation}
We call $\Pi(l_i)$ preferential splitting in this paper, because $\Pi (l_i)$ corresponds to the preferential attachment introduced in the original BA-model.
We chose the loop to be divided using random numbers and $\Pi(l_i)$. When the length of a daughter loop was smaller than the cutoff length $l_{\mathrm{min}}$, we removed the loop from the system. After $t$ time steps, the system has an assembly of loops from which we can get a VLD $n(l)$. By averaging $N_{\mathrm{samp}}$ samples, we obtained the ensemble average of VLD $\langle n(l)\rangle$.

\begin{figure}
\includegraphics[width=.9\linewidth]{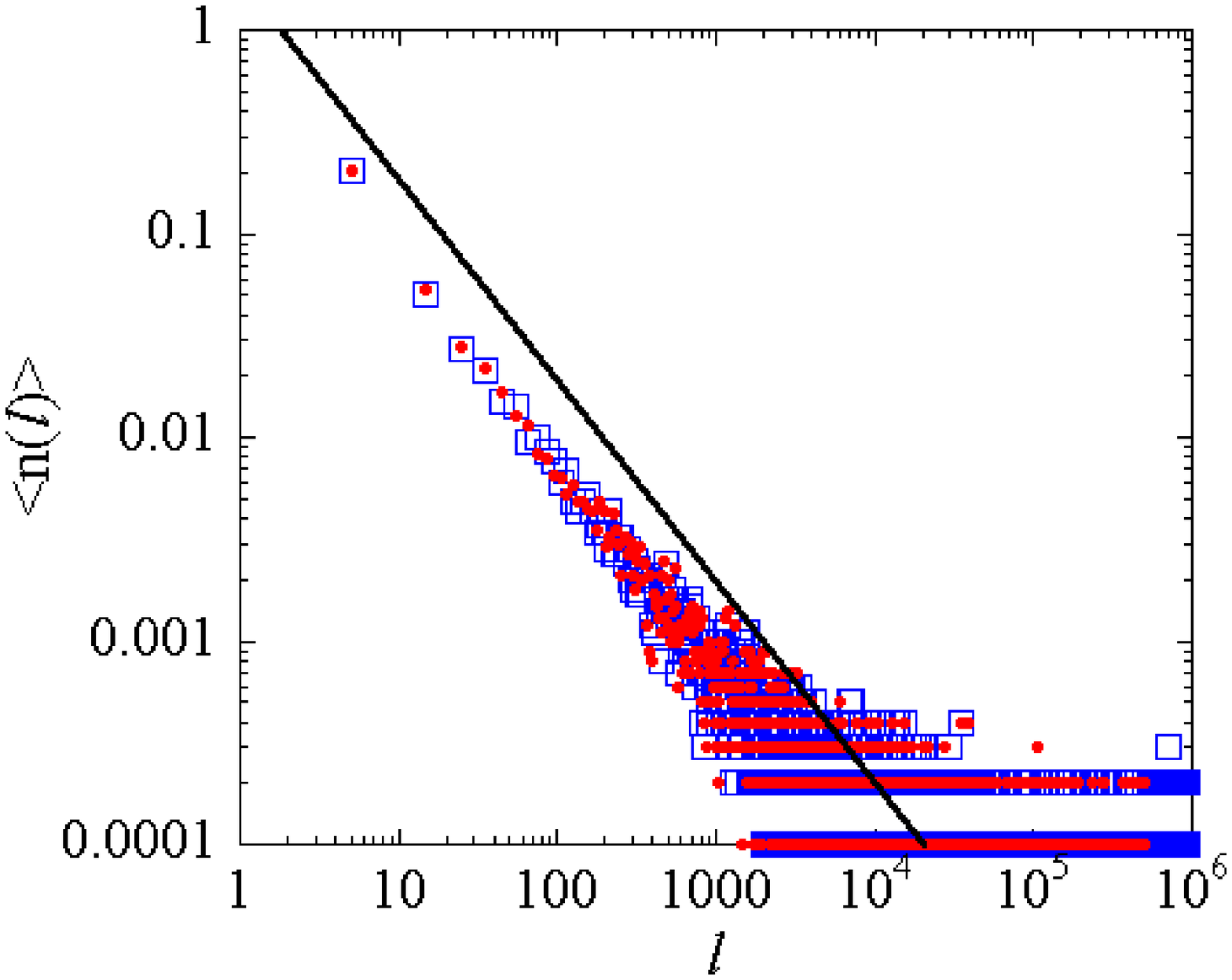}
\caption{Simulated $\langle n(l)\rangle$ at $t=100,000$ (blue open squares) and $t=1,000,000$(red closed circles). The simulations used $\Pi(l)$ from Eq.(\ref{eq:dep}). The slope of the solid line is given by the exponent $\alpha =1$.}
\label{fig:VLD}
\includegraphics[width=.9\linewidth]{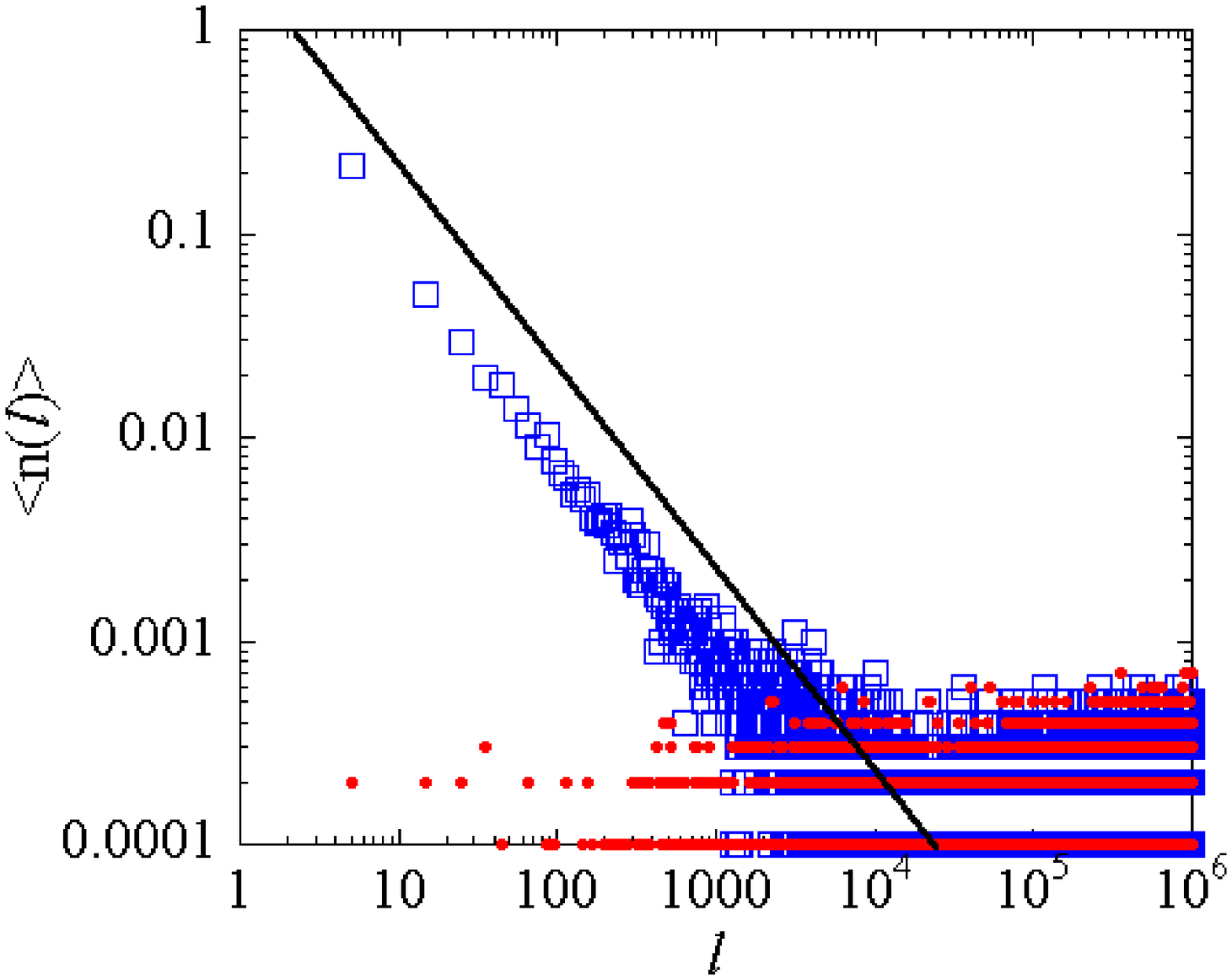}
\caption{Simulated $\langle n(l)\rangle$ at $t=100,000$ (blue open squares) starting from random $<n(l)>$(red closed circles) at $t=0$. The simulations used $\Pi(l)$ from Eq.(\ref{eq:dep}). The slope of the solid line is given by the exponent $\alpha =1$.}
\label{fig:rando_ini}
\end{figure}

Unlike the BA-model, which deals with growing networks, it is not easy to analytically derive the power exponent $\alpha$ of VLD, due mostly to the superlinear preferential splitting. Hence, we ran numerical simulations of the system. For the simulations, we assumed $m_0=1$, $N_{\mathrm{samp}}=1,000$, $L\equiv\Sigma_{j=1}^{m_0} l_j=10^6$, and $l_{\mathrm{min}}=1$. Figure \ref{fig:total_line} shows the resulting time dependence of the total line length. The total line length linearly decreases because the small-scale loops were removed. Our model did not reproduce the observed time dependence of total line length \cite{Tsubota2000} because the time variable in our model was the number of steps in the simulation. Figure \ref{fig:VLD} shows that our model system leads to a power law distribution of VLD. The obtained exponent $\alpha \approx 1$, which is consistent with the exponent obtained by Araki \textit{et.al.} and Kobayashi \textit{et.al.}.

\begin{figure}
\includegraphics[width=.9\linewidth]{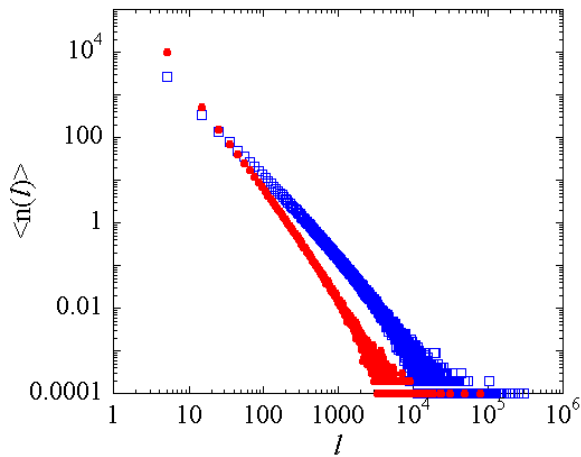}
\caption{Simulated $\langle n(l)\rangle$ at $t=100,000$ (blue open squares)  and $t=1,000,000$(red closed circles) for the case when $\Pi(l)$ came from Eq.(\ref{eq:const}) (i.e., constant preferential splitting). }
\label{fig:power_uni}
\end{figure}

 Because both the power law and the amplitude of the energy spectrum $E(k)$ hardly change during the decay \cite{Kobayashi2005}, we expect that a correct model should produce a distribution in which the main features are conserved during the decay. Figure \ref{fig:VLD} clearly demonstrates that VLD is independent of time especially in the small $l$ region, indicating that the system self-organizes its stationary scale-free VLD.

Figure \ref{fig:rando_ini} shows the obtained VLD that resulted from a random initial distribution of VLD. This shows that our model does produce a VLD with main features that are independent of the initial VLD; that is, this power law is very robust. We also confirmed that changing $l_{\mathrm{min}}$ hardly affects the power law of VLD as long as the ratio $L/l_{\mathrm{min}}$ is sufficiently large.

We now address whether or not the model results, particularly the stationary power law of VLD, required the preferential splitting $\Pi(l)$ of Eq. (\ref{eq:dep}). To examine this question, we investigated how VLD develops without the preferential splitting. Instead of Eq. (\ref{eq:dep}), we assumed
\begin{equation}
\Pi(l)=\frac{(1/l_i)^0}{\Sigma_j (1/l_j)^0}=\mathrm{const.}.
\label{eq:const}
\end{equation}
By running the same simulations, we found that the results lead to a similar power law; however, the exponent $\alpha$, grew monotonically with time (Fig.\ref{fig:power_uni}).
 
Our model reproduces the observed exponent well. To be accurate, however, our exponent is slightly smaller than the measured $\alpha \approx 1.38 \sim 1.5$ \cite{Araki2002, Kobayashi-private}. We do not understand the origin of this difference, but a likely explanation is the following. The dynamical scaling law, which was introduced through the preferential splitting, is a consequence of the localized induction approximation of vortex dynamics \cite{Schwarz1988}. Hence, we implicitly ignored the effect of the nonlocal interaction between vortices on the dynamical scaling. The nonlocal interaction may be needed to get good agreement with measurements. Or, we might need to include the combination-type reconnection in our model.

In conclusion, we have proposed an innovative model to study the origin of the power law of VLD. The nature of quantized vortices allowed us to describe the decay of quantum turbulence with a simple model that was similar to the Barab\'asi-Albert model of large networks. The two important generic mechanisms in the Barab\'asi-Albert model, the growth of networks and the preferential attachment, corresponded to the continuous self-reconnections and the preferential splitting due to the dynamical scaling law, respectively, in our model. Numerical simulation of our model resulted in a scale-invariant distribution of VLD. Furthermore, the exponent was consistent with measurements that found a VLD approximately proportional to $l^{-1}$. Our results suggest that only the Richardson cascade and a dynamical scaling law are needed to make the quantum turbulence self-organize with a scale-invariant VLD. This means that the emergence of power-law VLD is universal phenomenon in quantum turbulence beyond Kolmogorov quantum turbulence.

\begin{acknowledgments}
MT acknowledges support from a Grant-in-Aid for Scientific Research (Grant No. 15340122) by the Japan Society for the Promotion of Science.
\end{acknowledgments}



\end{document}